\documentclass[aps,prd,onecolumn,groupedaddress,showpacs,nofootinbib,amssymb]{revtex4}
\usepackage[dvips]{graphicx}
\usepackage{amssymb}
\usepackage{amsmath}
\usepackage{graphicx,color}
\usepackage{amsfonts}
\usepackage{bm}
\usepackage{cancel}
\usepackage{comment}
\newcommand{\be}{\begin{equation}}
\newcommand{\ee}{\end{equation}}
\newcommand{\bea}{\begin{eqnarray}}
\newcommand{\eea}{\end{eqnarray}}
\newcommand{\beaa}{\begin{eqnarray*}}
\newcommand{\eeaa}{\end{eqnarray*}}

\newcommand{\nn}{\nonumber \\}
\newcommand{\e}{\mathrm{e}}

\allowdisplaybreaks[4]
\begin{document}

\title{Propagation of Gravitational Waves in Chern-Simons Axion $F(R)$ Gravity}

\author{Shin'ichi~Nojiri,$^{1,2}$\,\thanks{nojiri@gravity.phys.nagoya-u.ac.jp}
S.D.~Odintsov,$^{3,4,5}$\,\thanks{odintsov@ieec.uab.es}
V.K.~Oikonomou,$^{6,7}$\,\thanks{v.k.oikonomou1979@gmail.com}
Arkady A. Popov,$^{8}$\,\thanks{arkady_popov@mail.ru}}

\affiliation{ $^{1)}$ Department of Physics, Nagoya University,
Nagoya 464-8602, Japan \\
$^{2)}$ Kobayashi-Maskawa Institute for the Origin of Particles
and the Universe, Nagoya University, Nagoya 464-8602, Japan \\
$^{3)}$ ICREA, Passeig Luis Companys, 23, 08010 Barcelona, Spain\\
$^{4)}$ Institute of Space Sciences (IEEC-CSIC) C. Can Magrans s/n,
08193 Barcelona, Spain\\
$^{5)}$ Tomsk State Pedagogical University, 634061 Tomsk, Russia\\
$^{6)}$ Department of Physics, Aristotle University of
Thessaloniki, Thessaloniki 54124, Greece\\
$^{7)}$ International Laboratory for Theoretical Cosmology, Tomsk
State University of Control Systems
and Radioelectronics (TUSUR), 634050 Tomsk, Russia\\
$^{8)}$ N. I. Lobachevsky Institute of Mathematics and Mechanics,
Kazan Federal University, 420008, Kremlevskaya street 18, Kazan,
Russia}


\begin{abstract}
In this paper we shall study the evolution of cosmological
gravitational waves in the context of Chern-Simons axion $F(R)$
gravity. In the case of Chern-Simons axion $F(R)$ gravity there
exist spin-0, spin-2 and spin-1 modes. As we demonstrate, from all
the gravitational waves modes of the Chern-Simons axion $F(R)$
gravity, only the two tensor modes are affected, while the spin-0
and spin-1 modes are not affected at all. With regard to the two
tensor modes, we show that these modes propagate in a
non-equivalent way, so the resulting tensor modes are chiral.
Notably, with regard to the propagation of the spin-2 graviton
modes, the structure of the dispersion relations becomes more
complicated in comparison with the Einstein gravity with the
Chern-Simons axion, but the resulting qualitative features of the
propagating modes are not changed. With regard to the spin-0 and
spin-1 modes, the Chern-Simons axion $F(R)$ gravity contains two
spin-0 modes and no vector spin-1 mode at all.  We also find that
for the very high energy mode, both the group velocity and the
phase velocity are proportional to the inverse of the square root
of the wave number, and therefore the velocities become smaller
for larger wave numbers or even vanish in the limit that the wave
number goes to infinity.
\end{abstract}


\maketitle

\section{Introduction}

The presence of a dark component of matter in the Universe was
assumed early after the first galactic rotation curves appeared.
Since then, theoretical physics studies were focused on proposing
massive particles weakly interacting with luminous matter, the
so-called WIMPs (Weakly Interacting Massive Particles), and there
exist examples coming from various theoretical contexts, see for
example
\cite{Bertone:2004pz,Mambrini:2015sia,Profumo:2013yn,Hooper:2007qk,Bergstrom:2000pn,Oikonomou:2006mh}.
To be honest, for the moment only indications exist that support
the particle nature of dark matter, such as the observational data
from the bullet cluster. However, nearly two decades of searches
did not result in finding any WIMP. A crucial point to stress is
that all the dark matter searches focused in mass ranges from a
few MeV up to hundreds of GeV's. However, only recently the
experimentalists focused their interest in searching WIMPs having
masses a few eV or even sub-eV.

String theory is up to date the most prominent theory that may
describe in a consistent way the quantum theory of gravity. One of
the interesting predictions of string theory is the existence of
low-mass axion
\cite{Marsh:2015xka,Sikivie:2006ni,Raffelt:2006cw,Linde:1991km}
particles, other than the QCD axions. The axions are particularly
appealing as dark matter candidates, since the mass range that
these may have, is not investigated yet, and only the last 5 years
experimentalists turned their focus on the axions. There is a
plethora of experimental
\cite{Du:2018uak,Henning:2018ogd,Ouellet:2018beu,Safdi:2018oeu,
Rozner:2019gba,Avignone:2018zpw,Caputo:2018vmy,Caputo:2018ljp,
Lawson:2019brd}
and theoretical proposals related to the axions
\cite{Marsh:2017yvc,Odintsov:2019mlf,Nojiri:2019nar,Nojiri:2019riz,
Odintsov:2019evb,Cicoli:2019ulk,Fukunaga:2019unq,
Caputo:2019joi,Sakharov:1994id,Sakharov:1996xg,Khlopov:1999tm,
Chang:2018rso,Irastorza:2018dyq,Anastassopoulos:2017ftl,
Sikivie:2014lha,Sikivie:2010bq,
Sikivie:2009qn,Caputo:2019tms,Masaki:2019ggg,Soda:2017sce,
Soda:2017dsu,Aoki:2017ehb,Masaki:2017aea,
Aoki:2016kwl,Obata:2016xcr,Aoki:2016mtn,Ikeda:2019fvj,
Arvanitaki:2019rax,Arvanitaki:2016qwi,Arvanitaki:2014wva,
Arvanitaki:2014dfa,Sen:2018cjt,Cardoso:2018tly,Rosa:2017ury,
Yoshino:2013ofa,Machado:2019xuc,Korochkin:2019qpe,Chou:2019enw,
Chang:2019tvx,Crisosto:2019fcj,Choi:2019jwx,Kavic:2019cgk,
Blas:2019qqp,Guerra:2019srj,Tenkanen:2019xzn,Huang:2019rmc,
Croon:2019iuh,Day:2019bbh,Kalashev:2018bra}.
Axions can induce particularly interesting effects in the
phenomenology of gravitational theories, since Chern-Simons terms
of the form $U(\phi)\tilde{R}R$ are allowed in the theory
\cite{Nishizawa:2018srh,Wagle:2018tyk,Yagi:2012vf,Yagi:2012ya,Molina:2010fb,
Izaurieta:2009hz,Sopuerta:2009iy,Konno:2009kg,Smith:2007jm,Matschull:1999he,
Haghani:2017yjk,Satoh:2007gn,Satoh:2008ck,Yoshida:2017cjl}. The
Chern-Simons terms produce non-equivalent polarizations in the
tensor modes of the underlying gravitational theory
\cite{Hwang:2005hb,Choi:1999zy}, and in the literature there exist
timely studies on chiral gravitational waves
\cite{Inomata:2018rin,Kamionkowski:1997av,Pritchard:2004qp,Lyth:2005jf,
Alexander:2007qe,Alexander:2004us,Cai:2016ihp,Moretti:2019yhs}.

In this paper we shall study a general Chern-Simons Axion $F(R)$
gravity
\cite{Nojiri:2017ncd,Nojiri:2010wj,Nojiri:2006ri,Capozziello:2011et,
Capozziello:2010zz,delaCruzDombriz:2012xy,Olmo:2011uz}, by mainly
focusing on the possibility of generating non-equivalent
polarizations in the tensor modes of the gravitational waves. We
shall be interested in the primordial gravity waves propagation,
assuming the conservation of the gravitational wave in the large
scale \cite{Choi:1999zy}. Our approach will mainly be
quantitative, since we will extract the gravitational wave
equations, and we shall study the tensor, vector and scalar modes.
As we demonstrate, the situation with the spin-0 scalar mode and
spin-1 vector mode, is not changed from the standard $F(R)$
gravity coupled with quintessence-like type scalar field and as we
demonstrate, the Chern-Simons axion term does not affect the
propagation of these modes at all. Particularly, there exist two
propagating scalar modes, and no propagating vector mode. With
regard to the two tensor modes, the resulting dispersion relations
are more complicated compared with the Einstein gravity with the
Chern-Simons axion term, but the qualitative features of the
propagating modes are not changed, and actually non-equivalent
propagation occur in both Einstein Chern-Simons and $F(R)$ gravity
Chern-Simons theories.

This paper is organized as follows: In section II we present the
general features of Chern-Simons Axion $F(R)$ gravity models. In
section III we extract the general gravitational wave equations,
while in section IV the various gravitational wave modes, and the
corresponding polarizations, are studied. Finally, the conclusions
follow in the end of the paper.

\section{The Chern-Simons Corrected Axion $F(R)$ Gravity}

In this paper we shall mainly consider a Axionic Chern-Simons
corrected $F(R)$ gravity model, whose action is given by,
\begin{equation}
\label{FCS5}
S = \frac{1}{2\kappa^2} \int d^4 x \sqrt{-g} \left[
F(R) - \frac{\omega(\phi)}{2}\partial_\mu \phi \partial^\mu \phi -
V(\phi) + U(\phi) \tilde\epsilon^{\mu\nu\rho\sigma} R^\tau_{\
\lambda\mu\nu} R^\lambda_{\ \tau\rho\sigma} \right] \, ,
\end{equation}
where we defined the totally antisymmetric Levi-Civita symbols
$\epsilon_{\mu\nu\rho\sigma}$ and $\epsilon^{\mu\nu\rho\sigma}$ as
follows,
\begin{equation}
\label{FCS2}
\epsilon_{0123} = - \epsilon^{0123} = 1 \, ,
\end{equation}
and in addition,
\begin{equation}
\label{FCS4B}
\epsilon_{\mu\nu\rho\sigma}
= \eta_{\mu\mu'} \eta_{\nu\nu'} \eta_{\rho\rho'} \eta_{\sigma\sigma'}
\epsilon^{\mu'\nu'\rho'\sigma'} \, .
\end{equation}
Then we obtain the following tensors,
\begin{equation}
\label{FCS4}
\tilde\epsilon^{\mu\nu\rho\sigma} \equiv
\frac{1}{\sqrt{-g}}\epsilon^{\mu\nu\rho\sigma} \, , \quad
\equiv g_{\mu\mu'} g_{\nu\nu'} g_{\rho\rho'} g_{\sigma\sigma'}
\tilde\epsilon^{\mu'\nu'\rho'\sigma'}
= \sqrt{-g} \epsilon_{\mu\nu\rho\sigma} \, .
\end{equation}
We should note that the following tensor identity holds true,
\begin{equation}
\label{FCS4D}
\nabla_\sigma \tilde\epsilon^{\zeta\eta\rho\xi} =0 \, .
\end{equation}
Since,
\begin{equation}
\label{FCS6}
\delta \left( \sqrt{-g} U(\phi) \tilde\epsilon^{\mu\nu\rho\sigma}
R^\tau_{\ \lambda\mu\nu}
R^\lambda_{\ \tau\rho\sigma} \right)
= 2 \sqrt{-g} U(\phi) \left[ \tilde\epsilon^{\zeta\eta\rho\mu}
R^{\tau\nu}_{\ \ \zeta\eta} + \tilde\epsilon^{\zeta\eta\rho\nu}
R^{\tau\mu}_{\ \ \zeta\eta} \right] \nabla_\rho \nabla_\tau
\delta g_{\mu\nu} \, ,
\end{equation}
upon varying the action (\ref{FCS5}) with respect to the metric,
we obtain the equations as follows,
\begin{align}
\label{FCS7}
0 =& \frac{1}{2}g_{\mu\nu} F(R) - R_{\mu\nu}F'(R) + \nabla_\mu \nabla_\nu F'(R)
 - g_{\mu\nu}\Box  F'(R) \nn
& + \frac{1}{2} \left\{ - \frac{\omega(\phi)}{2}\partial_\mu \phi \partial^\mu \phi
 - V(\phi) \right\} g_{\mu\nu}
+ \frac{\omega(\phi)}{2}\partial_\mu \phi \partial_\nu \phi \nn
& + 2 \left( g_{\mu\xi} g_{\nu\sigma} + g_{\mu\sigma} g_{\nu\xi} \right)
\nabla_\tau \nabla_\rho \left( U(\phi) \tilde\epsilon^{\zeta\eta\rho\xi}
R^{\tau\sigma}_{\ \ \zeta\eta} \right) \, .
\end{align}
The equation obtained from the variation of the action with
respect to the scalar field $\phi$ is given by,
\begin{equation}
\label{FCS5S}
0 = \nabla^\mu \left( \omega(\phi) \partial_\mu \phi \right) - V'(\phi)
+ U'(\phi) \epsilon^{\mu\nu\rho\sigma} R^\tau_{\ \lambda\mu\nu}
R^\lambda_{\ \tau\rho\sigma} \, .
\end{equation}
We now assume that the geometric background is a
Friedmann-Robertson-Walker (FRW) spacetime with flat spatial part,
\begin{equation}
\label{FRWmetric}
ds^2 = - dt^2 + a(t)^2 \sum_{i=1,2,3} \left( dx^i \right)^2 \, ,
\end{equation}
and we also assume that the scalar field $\phi$ depends solely on
the cosmological time $t$. In the FRW background, we obtain,
\begin{align}
\label{E2}
& \Gamma^t_{ij}=a^2 H \delta_{ij}\, ,\quad \Gamma^i_{jt}=\Gamma^i_{tj}=H\delta^i_{\ j}\, ,
\quad \Gamma^i_{jk}=\tilde \Gamma^i_{jk}\, ,\nn
& R_{itjt}=-\left(\dot H + H^2\right)a^2\delta_{ij}\, ,\quad
R_{ijkl}= a^4 H^2 \left(\delta_{ik} \delta_{lj}
 - \delta_{il} \delta_{kj}\right)\, ,\nn
& R_{tt}=-3\left(\dot H + H^2\right)\, ,\quad
R_{ij}=a^2 \left(\dot H + 3H^2\right)\delta_{ij}\, ,\quad
R= 6\dot H + 12 H^2\, , \nn
& \mbox{other components}=0\, ,
\end{align}
and from these we obtain the following equations,
\begin{align}
\label{FCS8}
0 =&  -\frac{1}{2}F(R_0) + 3\left(H^2  + \dot H\right) F'(R_0)
 - 18 \left(4H^2 \dot H + H \ddot H\right)F''(R_0)
+ \frac{\omega(\phi)}{4}{\dot\phi}^2 + \frac{V(\phi)}{2} \, , \nn
0=& \frac{1}{2}F(R_0) - \left(\dot H + 3H^2\right)F'(R_0)
+ 24 \left(4H^2 \dot H + {\dot H}^2 + 2 H \ddot H\right)F''(R_0)
+ 36\left(4H\dot H + \ddot H\right)^2F'''(R_0) \nn
& + \frac{\omega(\phi)}{4}{\dot\phi}^2 - \frac{V(\phi)}{2} \, .
\end{align}
Here $R_0=12H^2 + 6\dot H$ and we have neglected the contributions
from matter perfect fluids. We should note that the term
containing the scalar coupling function to the Chern-Simons term,
namely, $U(\phi)$, does not contribute to the above equations in
(\ref{FCS8}). We may choose $\phi$ to be the cosmological time
$t$, that is, $\phi=t$. Then the equations in (\ref{FCS8}) can be
rewritten as,
\begin{align}
\label{FCS9}
\omega(\phi) =& - 4 \dot H F'(R_0)
 - 4\left(12H^2 \dot H + 12 {\dot H}^2 + 15 H \ddot H\right)F''(R_0)
 - 72\left(4H\dot H + \ddot H\right)^2 F'''(R_0) \, , \nn
V(\phi)  =& F(R_0) - 2 \left(2 \dot H + 3H^2\right)F'(R_0)
+ \left(168 H^2 \dot H + 24 {\dot H}^2 + 66 H \ddot H\right)F''(R_0)
+ 36\left(4H\dot H + \ddot H\right)^2 F'''(R_0) \, .
\end{align}
Then if we choose,
\begin{align}
\label{FCS11}
\omega(\phi) =& - 4  f'(\phi) H F'(R_f)
 - 4\left(12f(\phi)^2 f'(\phi) + 12 f'(\phi)^2 + 15 f(\phi) f''(\phi) \right)F''(R_f) \nn
& - 72\left(4f(\phi) f'(\phi) + f''(\phi) \right)^2F'''(R_f) \, , \nn
V(\phi)  =& F(R_f) - 2 \left(2 f'(\phi) + 3 f(\phi)^2 \right)F'(R_f)
+ \left(168 f(\phi)^2 f'(\phi) + 24 f'(\phi)^2 + 66 f(\phi) f''(\phi) \right)F''(R_f) \nn
& + 36\left(4 f(\phi) f''(\phi) + f''(\phi)\right)^2F'''(R_f) \, , \nn
R_f \equiv& 12 f(\phi)^2 + 6 f'(\phi) \, .
\end{align}
a solution of the equations in (\ref{FCS8}) is given by $H=f(t)$ and $\phi=t$.

\section{Gravitational Wave Equations}

We now investigate the propagation of gravitational waves in the
Chern-Simons Axion $F(R)$ gravity. In order to study the
propagation of the gravitational waves, we consider the
perturbation of Eq.~(\ref{FCS7}), from the background whose metric
is $g^{(0)}_{\mu\nu}$,
\begin{equation}
\label{FCS14}
g_{\mu\nu} = g^{(0)}_{\mu\nu} + h_{\mu\nu} \, , \quad
\phi = \phi^{(0)} + \varphi \, .
\end{equation}
Then we can find the equations corresponding to the perturbed
Einstein equation, which are presented in the Appendix in Eq.
(\ref{FCS15}) due to the extended analytic form these have. On the
other hand, Eq.~(\ref{FCS5S}) gives,
\begin{align}
\label{FCS5Sp}
0 =& - \frac{1}{2} g^{(0)\mu\nu}h_{\mu\nu}\nabla^{(0)\rho}
\left( \omega\left(\phi^{(0)} \right) \partial_\rho \phi^{(0)} \right)
 - \nabla^{(0)}_\nu \left( h^{\nu\mu} \omega\left(\phi^{(0)} \right)
\partial_\mu \phi^{(0)} \right)
+ \frac{1}{2}\nabla^{(0)\rho} \left( g^{(0)\mu\nu}h_{\mu\nu}
\omega\left(\phi^{(0)} \right) \partial_\rho \phi^{(0)} \right) \nn
& + 2 U'\left(\phi^{(0)} \right) \left[ \tilde\epsilon^{(0)\zeta\eta\rho\mu}
R^{(0)\tau\nu}_{\ \ \ \ \ \zeta\eta} + \tilde\epsilon^{(0)\zeta\eta\rho\nu}
R^{(0) \tau\mu}_{\ \ \ \ \ \zeta\eta} \right] \nabla^{(0)}_\rho \nabla^{(0)}_\tau
h_{\mu\nu}
 - \frac{1}{2} g^{(0)\eta\zeta}h_{\eta\zeta} U'\left(\phi^{(0)} \right)
\tilde\epsilon^{(0) \mu\nu\rho\sigma} R^{(0)\tau}_{\ \ \ \ \lambda\mu\nu}
R^{(0)\lambda}_{\ \ \ \ \tau\rho\sigma} \nn
& + \nabla^{(0)\mu} \left( \omega'\left(\phi^{(0)} \right)
\varphi \partial_\mu \phi^{(0)} \right)
+ \nabla^{(0)\mu} \left( \omega\left(\phi^{(0)} \right) \partial_\mu \varphi \right)
 - V''\left(\phi^{(0)} \right) \varphi
+ U''\left(\phi^{(0)} \right) \varphi \tilde\epsilon^{(0) \mu\nu\rho\sigma}
R^{(0)\tau}_{\ \ \ \ \lambda\mu\nu}
R^{(0)\lambda}_{\ \ \ \ \tau\rho\sigma} \, .
\end{align}
The explicit form of the $(t,t)$, $(i,j)$, and $(t,i)$ components
of the modified Einstein equations $\delta G_{\mu \nu}=0$ from
Eq.~(\ref{FCS15}) in the FRW background (\ref{FRWmetric}) are
given as follows,
\begin{align}
\label{tt}
\delta G^t_{\ t}\equiv & g^{(0)t\mu} \delta G_{\mu t} \nn
=& F' \left\{-\left(\dot H +3 H^2 \right) h^t_{\ t}
+H \left( \partial_t h^i_{\ i} +2 \partial^i h^t_{\ i}  \right)
+\frac12 \partial^i \partial_j h^j_{\ i} -\frac12 \partial^i \partial_i h^j_{\ j} \right\} \nn
& + F'' \left\{ -6 \dddot H \ h^t_{\ t} +\ddot H \left[ H\left(-57 h^t_{\ t} -12 h^i_{\ i} \right)
 -3 \partial_t h^i_{\ i} -6\partial^i h^t_{\ i} \right] -18 {\dot H}^2  h^t_{\ t}
+\dot H \left[ H^2 \left( -30 h^t_{\ t} -48 h^i_{\ i} \right) \right. \right. \nn
& \left.
+H \left( -18 \partial_t h^t_{\ t} -12 \partial_t h^i_{\ i} -24 \partial^i h^t_{\ i} \right)
 -3 \partial^2_{t} h^i_{\ i} -6 \partial^i \partial_t h^t_{\ i} +9 \partial^i \partial_i h^t_{\ t}
 -3 \partial^i \partial_j h^j_{\ i} +3 \partial^i \partial_i h^j_{\ j} \right] \nn
& +36 H^4 h^t_{\ t}  +H^3  \left( -27 \partial_t h^t_{\ t} -12 \partial_t h^i_{\ i}
 -36 \partial^i h^t_{\ i} \right)
+H^2 \left( -9 \partial^2_{t} h^t_{\ t} + 9 \partial^2_t h^i_{\ i}
+21 \partial^i \partial_i h^t_{\ t} -6 \partial^i \partial_t h^t_{\ i} \right. \nn
& \left. -9 \partial^i \partial_j h^j_{\ i} +9 \partial^i \partial_i h^j_{\ j} \right)
+H \left( 3 \partial^3_{t} h^i_{\ i} +6 \partial^i \partial^2_{t} h^t_{\ i}
 -7\partial^i \partial_i \partial_t h^j_{\ j} +3 \partial^i \partial_j \partial_t h^j_{\ i}
 -4 \partial^i \partial^j  \partial_j  h^t_{\ i} \right) \nn
& \left. - \partial^i \partial_i \partial^2_{t} h^j_{\ j}
+ \partial^i \partial_i  \partial^k \partial_k h^t_{\ t}
 -2 \partial^j \partial_j  \partial^i \partial_t h^t_{\ i}
+ \partial^j \partial_j  \partial^k \partial_k h^i_{\ i}
- \partial^l \partial_l  \partial^j \partial^i h_{ij} \right\} \nn
& + F''' \left\{-54 {\ddot{H}}^2 h^t_{\ t} +\ddot{H} \left[ -540 \dot{H} H h^t_{\ t}
 -216 H^3 h^t_{\ t} + H^2 \left( -54 \partial_t h^t_{\ t} +72 \partial_t h^i_{\ i}
+72 \partial^i h^t_{\ i} \right) \right. \right. \nn
& \left. + H \left( 18 \partial^2_{t}  h^i_{\ i} +36 \partial^i \partial_t h^t_{\ i}
 -18 \partial^i \partial_i h^t_{\ t} +18 \partial^i \partial_j h^j_{\ i}
 -18 \partial^i \partial_i h^j_{\ j} \right) \right] -1296{\dot{H}}^2 H^2 h^t_{\ t} \nn
& +\dot{H} \left[ -864 H^4 h^t_{\ t} +H^3 \left( -216 \partial_t h^t_{\ t}
+288 \partial_t h^i_{\ i} +288 \partial^i h^t_{\ i} \right)
+ H^2 \left( 72 \partial^2_{t} h^i_{\ i} -72 \partial^i \partial_i h^t_{\ t} \right. \right. \nn
& \left. \left. +144 \partial^i \partial_t h^t_{\ i} +72  \partial^i \partial_j h^j_{\ i}
 -72  \partial^i \partial_i h^j_{\ j} \right) \right] \nn
& + \frac{1}{2} \left\{ \omega \left(\phi^{(0)} \right) \left( \dot\phi^{(0)}\right)^2
 - V \left( \phi^{(0)} \right) \right\} h^t_{\ t}
 - \frac{1}{2} \left\{ \frac{\omega' \left( \phi^{(0)} \right)}{2} \left( \dot\phi^{(0)}\right)^2 \varphi
+ \omega \left( \phi^{(0)} \right) \dot\phi^{(0)} \dot\varphi + V' \left( \phi^{(0)} \right) \varphi \right\} \nn
=& 0 \, ,
\end{align}
We define the Levi-Civita symbol, which is totally antisymmetric tensor,
in three dimensions as follows,
\begin{equation}
\label{LCsym}
\epsilon_{xyz}\equiv a^3 \, , \quad \epsilon^{ijk} \equiv g^{(0)il} g^{(0)jm} g^{(0)kn} \epsilon_{lmn}
= a^{-6} \epsilon_{ijk} \, .
\end{equation}
Then we can obtain the non-zero components of the perturbed
Einstein tensor. Specifically, the components $\delta G^i_{\ j}$
are presented in the Appendix in Eq. (\ref{Gij}) due to their
extended form. The corresponding $\delta G^t_{\ i}$ components
are,
\begin{align}
\label{Gti} \delta G^t_{\ i} \equiv & g^{(0)t\mu} \delta G_{\mu i}
\nn =& \frac{2 \dot U}{a^2} \epsilon_{ikl} \left( \partial^k
\partial^m \partial_t h^l_{\ m} + \partial^m \partial_m \partial^k
h^{tl} \right)  \nn & +F' \left\{  H \partial_i h^t_{\ t} +
\frac{1}{2} \partial_i \partial^t h^k_{\ k}
 - \frac{1}{2} \partial^k \partial^t h_{ik}
+\frac{1}{2} \partial^k \partial_k h^t_{\ i}
- \frac{1}{2} \partial^k \partial_i h^t_k \right\} \nn
& +F'' \left\{ \ddot{H} \left( -42H h^t_{\ i} + 3\partial_i h^t_{\ t} \right)
 -24{\dot{H}}^2 h^t_{\ i}
+\dot{H}\left[ -72H^2 h^t_{\ i} +6H \partial_i h^t_{\ t}
+\partial_i \partial_t \left( 9h^t_{\ t} -4 h^k_{\ k} \right) \right. \right. \nn
& \left. - 4 \partial_i \partial^k h^t_{\ k} \right] -12H^3 \partial_i h^t_{\ t}
+H^2 \left[ \partial_i \partial_t \left( 9h^t_{\ t} +4h^k_{\ k} \right)
+ 12 \partial_i \partial^k h^t_{\ x} \right]  \nn
& +H\left[ 3\partial_i \partial_t^2 \left( h^t_{\ t} -h^k_{\ k} \right)
 -3\partial_i \partial^k \partial_k \left(h^t_{\ t} + h^k_{\ k} \right)
+ 3\partial_i \partial^k \partial^l h_{kl}
+ 2 \partial_i \partial^k \partial_t h^t_{\ k} \right] \nn
& \left. - \partial_i \partial_t^3 h^k_{\ k}
+ \partial_i \partial^k \partial_k \partial_t \left( h^t_{\ t} + h^l_{\ l} \right)
 - \partial_i \partial^k \partial^l \partial_t h_{kl}
 - 2 \partial_i \partial^k \partial_t^2 h^t_{\ k} \right\} \nn
& + F''' \left\{ -36{\ddot{H}}^2 h^t_{\ i}
+ \ddot{H} \left[ \dot{H} \left( -288 H h^t_{\ i}
+36\partial_i h^t_{\ t} \right) + 72 H^2 \partial_i h^t_{\ t}
+H\partial_i \partial_t \left( 18h^t_{\ t} -24 h^k_{\ k} \right) \right. \right. \nn
& \left.  -24H \partial_i \partial^k h^t_{\ k}
 - 6 \partial_i \partial_t^2 h^k_{\ k}
+  6\partial_i \partial^k \partial_k \left(h^t_{\ t} +h^l_{\ l} \right)
 - 6 \partial_i \partial^k \partial^l h_{kl}
 - 12 \partial_i \partial^k \partial_t h^t_{\ k} \right] \nn
& +{\dot{H}}^2 \left( -576 H^2 h^t_{\ i} +144H \partial_i h^t_{\ t} \right)
+\dot{H} \left[  288H^3 \partial_i h^t_{\ t}
+ H^2\left( \partial_i \partial_t \left(72h^t_{\ t} -96 h^k_{\ k} \right)
\right. \right. \nn
& \left. \left. - 96 \partial_i \partial^k h^t_{\ k}
-24 H \partial_i \partial_t^2 h^k_{\ k}
+ H \left( 24 \partial_i \partial^k \partial_k \left( h^t_{\ t} + h^l_{\ l} \right)
 - 24 \partial_i \partial^k \partial^l h_{kl}
 - 48\partial_i \partial^k h^t_{\ k} \right) \right] \right\} \nn
& + \frac{1}{2} \left\{ - \frac{\omega \left(\phi^{(0)}\right)}{2}
\partial_\mu \phi^{(0)} \partial^\mu \phi^{(0)} - V \left( \phi^{(0)} \right) \right\} h^t_{\ i}
 - \frac{\omega \left(\phi^{(0)}  \right)}{2}\dot{\phi^{(0)}} \partial_i \varphi \nn
=& 0\, .
\end{align}
We should note that all the terms coming from the last term
in Eq.~(\ref{FCS15}) identically vanish. A more explicit form of
Eq.~(\ref{FCS5Sp}) is given by,
\begin{align}
\label{FCS5SpEx}
0 =& \frac{1}{2} g^{(0)\mu\nu}h_{\mu\nu}
\left( \partial_t + 3 H \right) \left( \omega\left(\phi^{(0)} \right) \dot\phi^{(0)} \right)
 - \partial_\nu \left( h^{\nu t} \omega\left(\phi^{(0)} \right) \dot \phi^{(0)} \right)
 - 3H h^{tt} \partial_t \left( \omega\left(\phi^{(0)} \right) \dot \phi^{(0)} \right) \nn
& - \frac{1}{2} \left( \partial_t + 3 H \right) \left( g^{(0)\mu\nu}h_{\mu\nu}
\omega\left(\phi^{(0)} \right) \dot\phi^{(0)} \right) \nn
& - \left( \partial_t + 3 H \right) \left( \omega'\left(\phi^{(0)} \right)
\varphi \dot\phi^{(0)} \right)
 - \left( \partial_t + 3 H \right) \left( \omega\left(\phi^{(0)} \right) \dot\varphi \right)
+ a^3 \omega\left(\phi^{(0)} \right) \partial^k \partial_k \varphi
 - V''\left(\phi^{(0)} \right) \varphi \, .
\end{align}

\section{Polarizations of Gravitational Waves}

The most important study in the Chern-Simons Axion $F(R)$ gravity
is related to the polarization modes of the gravitational waves.
We now consider the following modes,
\begin{itemize}
\item Spin 2 tensor mode
\begin{equation}
\label{spin2}
\hat h_{ij}\, , \quad
h_{it}=h_{tt}=0\, , \quad h^i_{\ i} =0 \, ,
\quad \partial^j h_{ij} = 0 \, , \quad
\left(i=x,y,z\right) \ , \quad \varphi=0 \, .
\end{equation}
\item Spin 1 vector mode
\begin{equation}
\label{spin1}
A_i \equiv \partial^j h_{ji} \, , \quad \partial^i \partial^j h_{ij} = \partial^i A_i =0 \, ,
\quad h_{it}=h_{tt}=0\, , \quad h^i_{\ i} =0 \, , \quad \left(i,j=x,y,z\right) \ , \quad \varphi=0 \, .
\end{equation}
\item Spin 0 scalar mode
\begin{equation}
\label{spin0}
h = h^i_{\ i} \, . \quad h_{ij} =\partial_i \partial_j B
 - \frac{1}{3} g^{(0)}_{ij} \partial^k \partial_k B \, , \quad
\varphi \, .
\end{equation}
\end{itemize}
Then  the tensor $h_{ij}$ can be decomposed as follows,
\begin{equation}
\label{decompose}
h_{ij} = \hat h_{ij} + \frac{1}{2} \left( \partial^k \partial_k \right)^{-1}
\left( \partial_i A_j + \partial_j A_i \right) + \frac{1}{3} g^{(0)}_{ij} h
+ \partial_i \partial_j B  - \frac{1}{3} g^{(0)}_{ij} \partial^k \partial_k B \, ,
\end{equation}
which gives,
\begin{equation}
\label{B}
B = \frac{3}{2} \left( \partial^k \partial_k \right)^{-2} \partial^i \partial^j h_{ij}
 - \frac{1}{2} \left( \partial^k \partial_k \right)^{-1} h \quad \mbox{or} \quad
\partial^i \partial^j h_{ij} = \frac{2}{3} \left( \partial^k \partial_k \right)^2 B
+ \frac{1}{3} \partial^k \partial_k h \, .
\end{equation}
For all the above modes we considered, we have implicitly chosen
the gauge condition $h_{t\mu}=0$.

\subsection{Spin 2 tensor mode}

Let us study in some detail the spin 2 tensor mode. In the case of
spin 2 mode, we find that $\delta G^t_{\ t}$ in (\ref{tt}) and
$\delta G^t_{\ i}$ in Eq.~(\ref{Gti}) trivially vanish. In
addition $\delta G^t_{\ t}=\delta G^t_{\ i}=0$ and
Eq.~(\ref{FCS5SpEx}) is also trivially satisfied. On the other
hand, $\delta G^i_{\ j}$ in (\ref{Gij}) has the following form,
\begin{align}
\label{GijS2}
\delta G^i_{\ j}=& - \epsilon^{klm}\left( \delta^i_{\ k} g^{(0)}_{jn}
+ \delta^i_{\ n} g^{(0)}_{jk} \right)
\left\{ \left( 4H \dot{U} +  2 \ddot{U} \right) \partial_l \partial_t \hat h^n_{\ m}
+ 2 \dot{U} \partial_l \partial_t^2 \hat h^n_{\ m} \right\} \nn
& +F' \left\{ - \frac{3}{2} H \partial_t \hat h^i_{\ j}
 - \frac{1}{2} \partial_t^2 \hat h^i_{\ j} + \frac{1}{2} \partial^k \partial_k \hat h^i_{\ j} \right\} \nn
& +F'' \left\{ \ddot{H} \left[  -42 H \hat h^i_{\ j} + 3 \partial_{t} \hat h^i_{\ j} \right]
 - 24 {\dot{H}}^2 \hat h^i_{\ j}
+ \dot{H} \left[ - 72 H^2 \hat h^i_{\ j} + 12 \partial_{t} \hat h^i_{\ j} \right] \right\} \nn
& +F''' \left\{ - 36 {\ddot{H}}^2 \hat h^i_{\ j}
 - 288 \ddot{H} \dot{H} H \hat h^i_{\ j} - 576 {\dot{H}}^2 H^2 \hat h^i_{\ j} \right\}
+ \frac{1}{2} \left\{ \frac{\omega \left(\phi^{(0)}\right)}{2} \left( \dot\phi^{(0)} \right)^2
 - V \left( \phi^{(0)} \right) \right\} {\hat h}^i_{\ j} \, .
\end{align}
We should note that the obtained equation (\ref{GijS2}) is a
second order differential equation with respect to the cosmic time
$t$, although the original equation (\ref{FCS7}) or (\ref{Gij}) is
a fourth order difference equation. This is not curious, because
the $F(R)$ gravity action can be rewritten in the form of a
scalar-tensor theory, that is, the rewritten action is the sum of
the Einstein-Hilbert action and the action of the scalar field
with potential. So in effect, the spin-two mode originates from
the Einstein-Hilbert part, which gives the standard Einstein
equation, that is, the second order differential equation. The
existence of $U$-terms in (\ref{GijS2}) give the mixing of
$+$-mode and $x$-mode and the dispersion relation of the
left-polarized mode is different from that of the right-polarized
mode (see also \cite{Seto:2008sr}, for example). We should note
that there appear terms including the first derivative with
respect to the cosmic time, $\partial_{t} \hat h^i_{\ j}$, which
generate an enhancement or dissipation of the gravitational wave.
By using (\ref{FCS8}), we may rewrite (\ref{GijS2}) as follows,
\begin{align}
\label{GijS3}
\delta G^i_{\ j} =& - \epsilon^{klm}\left( \delta^i_{\ k} g^{(0)}_{jn}
+ \delta^i_{\ n} g^{(0)}_{jk} \right)
\left\{ \left( 4H \dot{U} +  2 \ddot{U} \right) \partial_l \partial_t \hat h^n_{\ m}
+ 2 \dot{U} \partial_l \partial_t^2 \hat h^n_{\ m} \right\} \nn
& - \frac{1}{2}F \hat h^i_{\ j} + F' \left\{ - \frac{3}{2} H \partial_t \hat h^i_{\ j}
 - \frac{1}{2} \partial_t^2 \hat h^i_{\ j} + \frac{1}{2} \partial^k \partial_k \hat h^i_{\ j}
+ \left(\dot H + 3H^2\right) \hat h^i_{\ j} \right\} \nn
& +F'' \left\{ 3 \ddot{H} \partial_{t} \hat h^i_{\ j}
+ 12 \dot{H} 2 \partial_{t} \hat h^i_{\ j}
 - \left( 42 H \ddot{H} + 168 H^2 \dot H + 48 {\dot H}^2 + 48 H \ddot H\right)
\hat h^i_{\ j}
\right\} \nn
& +F''' \left\{ - 72 {\ddot{H}}^2
 - 576 \ddot{H} \dot{H} H - 1152 {\dot{H}}^2 H^2 \right\} \hat h^i_{\ j} \, ,
\end{align}
Just for simplicity, we consider the case that $\dot U$ and $H$,
and therefore, $F$, $F'$, $F''$, and $F'''$ can be regarded as
constants. Then Eq.~(\ref{GijS3}) can be reduced as follows,
\begin{align}
\label{GijS3B}
\delta G^i_{\ j} =& - \epsilon^{klm}\left( \delta^i_{\ k} g^{(0)}_{jn}
+ \delta^i_{\ n} g^{(0)}_{jk} \right) \dot{U} \left\{ 4H \partial_l \partial_t \hat h^n_{\ m}
+ 2 \partial_l \partial_t^2 \hat h^n_{\ m} \right\} \nn
& - \frac{1}{2}F \hat h^i_{\ j} + F' \left\{ - \frac{3}{2} H \partial_t \hat h^i_{\ j}
 - \frac{1}{2} \partial_t^2 \hat h^i_{\ j} + \frac{1}{2} \partial^k \partial_k \hat h^i_{\ j}
+ 3H^2 \hat h^i_{\ j} \right\} \, ,
\end{align}
We consider the plane wave propagating in the $z$-direction with
the wave number $k$ and frequency $\omega$, $\hat h^i_{\ j} =
h^{(0)i}_{\ \ \ \ j} \e^{-i\omega t + ikz}$ with constants
$h^{(0)i}_{\ \ \ \ j}$. Then Eq.~(\ref{spin2}) tells $h^{(0)z}_{\
\ \ \ j}=h^{(0)i}_{\ \ \ \ z}=0$ and in effect we find,
\begin{align}
\label{GijS4}
\delta G^x_{\ x} =& 2 \omega k \dot{U} \left\{ 4H - 2i\omega \right\} h^{(0)x}_{\ \ \ \ y}
+ \left[ - \frac{1}{2}F + F' \left\{ \frac{3}{2} i \omega H
+ \frac{1}{2} \omega^2 - \frac{1}{2} k^2
+ 3H^2 \right\} \right] h^{(0)x}_{\ \ \ \ x}\, , \nn
\delta G^y_{\ y} =& -2 \omega k \dot{U} \left\{ 4H - 2i\omega \right\} h^{(0)y}_{\ \ \ \ x}
+ \left[ - \frac{1}{2}F + F' \left\{ \frac{3}{2} i \omega H
+ \frac{1}{2} \omega^2 - \frac{1}{2} k^2
+ 3H^2 \right\} \right] h^{(0)y}_{\ \ \ \ y}\, , \nn
\delta G^x_{\ y} =& 2 \omega k \dot{U} \left\{ 4H - 2i\omega \right\} h^{(0)y}_{\ \ \ \ y}
+ \left[ - \frac{1}{2}F + F' \left\{ \frac{3}{2} i \omega H
+ \frac{1}{2} \omega^2 - \frac{1}{2} k^2
+ 3H^2 \right\} \right] h^{(0)x}_{\ \ \ \ y}\, , \nn
\delta G^y_{\ x} =& - 2 \omega k \dot{U} \left\{ 4H - 2i\omega \right\}
h^{(0)x}_{\ \ \ \ x}
+ \left[ - \frac{1}{2}F + F' \left\{ \frac{3}{2} i \omega H
+ \frac{1}{2} \omega^2 - \frac{1}{2} k^2
+ 3H^2 \right\} \right] h^{(0)y}_{\ \ \ \ x}\, , \nn
\delta G^z_{\ i} =&\delta G^i_{\ z} = 0 \, .
\end{align}
Usually we consider the following two modes, namely the $+$ mode
where $h^{(0)}_+ \equiv h^{(0)x}_{\ \ \ \ x}= - h^{(0)y}_{\ \ \ \
y}$ and the $\times$ mode where $h^{(0)}_\times \equiv h^{(0)x}_{\
\ \ \ y}= h^{(0)y}_{\ \ \ \ x}$. In terms of this mode, the
non-trivial equations in (\ref{GijS4}) take the following forms,
\begin{align}
\label{GijS5}
0 =& 2 \omega k \dot{U} \left\{ 4H - 2i\omega \right\} h^{(0)}_\times
+ \left[ - \frac{1}{2}F + F' \left\{ \frac{3}{2} i \omega H
+ \frac{1}{2} \omega^2 - \frac{1}{2} k^2
+ 3H^2 \right\} \right] h^{(0)}_+ \, , \nn
0 =& - 2 \omega k \dot{U} \left\{ 4H - 2i\omega \right\} h^{(0)}_+
+ \left[ - \frac{1}{2}F + F' \left\{ \frac{3}{2} i \omega H
+ \frac{1}{2} \omega^2 - \frac{1}{2} k^2
+ 3H^2 \right\} \right] h^{(0)}_\times\, .
\end{align}
The above equations indicate that there should always be a mixing
between the $+$ mode and $\times$ mode and they should appear in
the forms of the left-handed or right handed mode, where
$h^{(0)}_+=\pm h^{(0)}_\times$. The equations in (\ref{GijS5})
give also the following dispersion relation,
\begin{equation}
\label{GijS6}
0 = \pm 2 \omega k \dot{U} \left\{ 4H - 2i\omega \right\}
 - \frac{1}{2}F + F' \left\{ \frac{3}{2} i \omega H
+ \frac{1}{2} \omega^2 - \frac{1}{2} k^2
+ 3H^2 \right\} \ \, .
\end{equation}
The terms including $i\omega$ in (\ref{GijS6}) come from the terms including
$\partial_{t} \hat h^i_{\ j}$ in (\ref{GijS2}), which generate the
enhancement or dissipation of the gravitational wave.
In the case of the Chern-Simons axion Einstein gravity \cite{Nojiri:2019nar},
we have $F'=1$ and $F=R\sim 24 H^2$. Therefore the qualitative
structure of the dispersion relation and the left- and
right-handed modes are not so changed from those corresponding to
the Chern-Simons axion Einstein gravity. In case of the
Chern-Simons axion $F(R)$ gravity, $F(R)$ depends on the time
coordinate $t$ and therefore the full solution of the
gravitational wave in (\ref{GijS3B}) becomes rather complicated. We
should note that the polarization of the gravitational wave in the
early Universe also affects the polarization of CMB, and
specifically the E-mode and B-modes, see for example
\cite{Bielefeld:2014nza}.

We now investigate the dispersion relation (\ref{GijS5}) in more
detail. Eq.~(\ref{GijS5}) can be solved as,
\begin{equation}
\label{sol1}
\omega =  \frac{ - \left( \frac{3}{2}i F' H +  \delta_\mathrm{LR} 8k \dot U H \right)
\pm \sqrt{ {F'}^2 k^2 - \frac{33}{4} {F'}^2 H^2 + 64 k^2 {\dot U}^2 H^2 + F' F
+  i \delta_\mathrm{LR} \left( 72 k F' \dot U H^2 - 8k \dot U F
 - 8k \dot U F' k^3 \right)}}
{2 \left( \frac{F'}{2} -  \delta_\mathrm{LR} 4ik \dot U \right) } \, .
\end{equation}
Here $ \delta_\mathrm{LR}=\pm 1$ for left or right-handed mode.
For the high energy mode, for which $k\gg H$ and by neglecting the
contribution from the Chern-Simons term, that is, $k\ll
\frac{F'}{\dot U}$, we obtain $\omega \sim \pm k$. Therefore the
propagating speed of the gravitational wave is not changed. When
$\omega$ is real number, $\omega$ should be positive and we should
choose the plus sign $+$ of $\pm$ in (\ref{sol1}). We should note,
however, that for the very high energy mode when $\dot U$ does not
vanish, $\dot U \neq 0$, that is, the mode $k\gg H$ and $k \gg
\frac{F'}{\dot U}$, Eq.~(\ref{GijS6}) has the form $0 = -
\delta_\mathrm{LR} 4i \omega^2 k \dot{U}
 - \frac{F'}{2} k^2$ and therefore we find
$\omega^2 \sim - \delta_\mathrm{LR} i \frac{F'}{8U'} k$, which is
rather strange dispersion relation. By assuming that the real part
of $\omega$ is positive we obtain $\omega = \e^{\pm \frac{\pi}{4}}
\sqrt{ \left| \frac{F'}{8U'} k \right|}$. Therefore there always
appear an amplified and a decaying mode. Furthermore the group
velocity $v_\mathrm{g}$ and the phase velocity $v_\mathrm{p}$ are
given by $v_\mathrm{g} \equiv \frac{d\omega}{dk} \propto
\frac{1}{\sqrt{k}}$ and $v_\mathrm{p} \equiv \frac{\omega}{k}
\propto \frac{1}{\sqrt{k}}$, which become smaller for larger $k$
and much smaller than the velocity of light.

\subsection{Spin 1 Vector Mode}

In the case of the spin 1 mode, we find $\delta G^t_{\ t}$ in
(\ref{tt}) vanishes and Eq.~(\ref{FCS5SpEx}) is also trivially
satisfied, again, $\delta G^t_{\ t}=0$. On the other hand, $\delta
G^t_{\ i}$ in (\ref{Gti}) has the following form,
\begin{equation}
\label{GtiV1}
\delta G^t_{\ i}= \frac{2 \dot U}{a^2} \epsilon_{ikl} \partial^k \partial_t A^l
 - \frac{1}{2} F' \partial^t A_i =0 \, ,
\end{equation}
and $\delta G^i_{\ j}$ in Eq.~(\ref{Gij}) has the following form,
\begin{align}
\label{GijV1}
\delta G^i_{\ j}
=& \epsilon^{klm}\left( \delta^i_{\ k} g^{(0)}_{jn} + \delta^i_{\ n} g^{(0)}_{jk} \right)
\left[ - \left\{ \left( 4H \dot{U} +  2 \ddot{U} \right)
\partial_l \partial_t {h^{(A)}}^n_{\ m}
+ 2 \dot{U} \partial_l \partial_t^2 {h^{(A)}}^n_{\ m} \right\}
 - 2 \dot{U} \partial_l \partial^n A_m \right] \nn
& +F' \left\{ - \frac{3}{2} H \partial_t {h^{(A)}}^i_{\ j}
 - \frac{1}{2} \partial^2_t {h^{(A)}}^i_{\ j}
+ \frac{1}{2} \partial^k \partial_k {h^{(A)}}^i_{\ j}
 - \frac{1}{2} \left( \partial^i A_j + g^{(0)im} g^{(0)}_{jl} \partial^l A_m \right) \right\}  \nn
& +F'' \left\{ \ddot{H} \left[  -42 H {h^{(A)}}^i_{\ j} + 3 \partial_{t} {h^{(A)}}^i_{\ j} \right]
 - 24 {\dot{H}}^2 {h^{(A)}}^i_{\ j}
+ \dot{H} \left[ - 72 H^2 {h^{(A)}}^i_{\ j}
+ 12 \partial_{t} {h^{(A)}}^i_{\ j} \right] \right\} \nn
& +F''' \left\{ - 36 {\ddot{H}}^2 {h^{(A)}}^i_{\ j}
 - 288 \ddot{H} \dot{H} H {h^{(A)}}^i_{\ j}
 - 576 {\dot{H}}^2 H^2 {h^{(A)}}^i_{\ j} \right\}
+ \frac{1}{2} \left\{ \frac{\omega \left(\phi^{(0)}\right)}{2} \left( \dot\phi^{(0)} \right)^2
 - V \left( \phi^{(0)} \right) \right\} {h^{(A)}}^i_{\ j}  \nn
=& 0 \, ,
\end{align}
where $h^{(A)}_{ij}$ is,
\begin{equation}
\label{hA}
h^{(A)}_{ij} \equiv \frac{1}{2} \left( \partial^k \partial_k \right)^{-1}
\left( \partial_i A_j + \partial_j A_i \right) \, ,
\end{equation}
We now discuss the qualitative implications of Eq.~(\ref{GtiV1}).
When $U=0$ as in the standard $F(R)$ gravity, we obtain $\dot A_i
=0$. In effect, there is no time evolution of $A_i$, or no
propagating mode and therefore $A_i$ is determined by the initial
conditions consistent with (\ref{GijV1}). When $U\neq 0$, since
there is a rotational symmetry, we may consider the plane wave
propagating in the $z$-direction with the wave number $k$, $A_i =
\alpha_i(t) \e^{ikz}$. Eq.~(\ref{spin1}) also indicates that $A_z
= \alpha_z(t)=0$. Then the $i=z$ component in Eq.~(\ref{GtiV1}) is
trivially satisfied and we obtain the following non-trivial
equations,
\begin{equation}
\label{V1}
0 = - \frac{2i k \dot U}{a} \partial_t \left( a^{-2} \alpha_y (t) \right)
+ \frac{1}{2} F' \dot\alpha_x (t) \, , \quad
0 = \frac{2i k \dot U}{a} \partial_t \left( a^{-2} \alpha_x (t) \right)
+ \frac{1}{2} F' \dot\alpha_y (t) \, ,
\end{equation}
which can be rewritten in a matrix form as follows,
\begin{equation}
\label{V2}
\left( \begin{array}{cc}
\frac{2i k \dot U}{a^3} & \frac{1}{2} F' \\
\frac{1}{2} F' & - \frac{2i k \dot U}{a^3}
\end{array} \right)
\left( \begin{array}{c}
\dot\alpha_x \\
\dot\alpha_y
\end{array} \right)
= \frac{2i k \dot U H}{a^3}
\left( \begin{array}{c}
 - \alpha_x \\
\alpha_y
\end{array} \right) \, .
\end{equation}
As an example, we consider the case that $H$, $\frac{\dot U}{a^3}$
and $F'$ are constant. Then by assuming $\alpha_x= \alpha_x^{(0)}
\e^{-i\omega t}$ and $\alpha_y= \alpha_y^{(0)} \e^{-i\omega t}$
with constants $\omega$, $\alpha_x^{(0)}$ and $\alpha_y^{(0)}$, we
obtain,
\begin{equation}
\label{V3}
\left( \begin{array}{cc}
\frac{2i k \dot U}{a^3} \left( - i \omega + H \right)
& \frac{1}{2} F' \\
\frac{1}{2} F' & - \frac{2i k \dot U}{a^3} \left( - i \omega - H \right)
\end{array} \right)
\left( \begin{array}{c}
\alpha_x^{(0)} \\
\alpha_y^{(0)}
\end{array} \right)
= 0 \, .
\end{equation}
In order that Eq.~(\ref{V3}) has non-trivial solutions for
$\alpha_x^{(0)}$ and $\alpha_y^{(0)}$, the determinant of the
matrix should vanish, and this constraint gives the following
dispersion relation,
\begin{equation}
\label{V4}
0 = \frac{4 k^2 {\dot U}^2}{a^6} \left( \omega^2 + H^2 \right)
+ \frac{1}{4} {F'}^2 \, ,
\end{equation}
which indicates that $\omega$ must be purely imaginary and
therefore there is no propagating mode. The above result is true
for the high frequency mode, even if $H$, $\frac{\dot U}{a^3}$ and
$F'$ are not constant. Therefore, there is no propagating mode of
Spin 1, which is consistent with the standard requirement coming
from the general covariance.

\subsection{Spin 0 scalar mode}

Let us now consider the spin-0 mode, which is also present in the
pure $F(R)$ gravity. For the spin-0 mode, we find,
\begin{align}
\label{ttS}
\delta G^t_{\ t}=& \frac{1}{3} F' \left\{ \left(\partial^k \partial_k \right)^2 B  - \partial^k \partial_k h \right\} \nn
& + F'' \left\{ \ddot H \left[ -12 H h -3 \partial_t h \right]
+ \dot H \left[ - 48 H^2 h - 12 H \partial_t h  -3 \partial_t^2 h
 - 2 \left( \partial^k \partial_k \right)^2 B + 2 \partial^k \partial_k h \right] \right. \nn
& - 12 H^3 \partial_t h + H^2 \left( 9 \partial^2_t h - 6 \left( \partial^k \partial_k \right)^2 B
+ 6 \partial^k \partial_k h \right)
+ H \left( 3 \partial_t^3 h + \left( \partial_t + 2 H \right)
\left( 2 \left( \partial^k \partial_k \right)^2 B - 6 \partial^k \partial_k h \right) \right) \nn
& \left. - \partial^i \partial_i \partial^2_{t} h
- \frac{2}{3} \partial^l \partial_l \left( \left( \partial^k \partial_k \right)^2 B - \partial^k \partial_k h \right) \right\} \nn
& + F''' \left\{ \ddot{H} \left[ 72 H^2 \partial_t h + 18 H \left( \partial^2_{t}  h
+12 \left( \left( \partial^k \partial_k \right)^2 B - \partial^k \partial_k h \right) \right) \right] \right. \nn
& +\dot{H} \left[ 288 H^3 \partial_t h
+ 72 H^2 \left( \partial^2_{t} h
+ \left( \frac{2}{3} \left( \partial^k \partial_k \right)^2 B + \frac{1}{3} \partial^k \partial_k h \right)
 - \partial^i \partial_i h \right) \right] \nn
& - \frac{1}{2} \left\{ \frac{\omega' \left( \phi^{(0)} \right)}{2} \left( \dot\phi^{(0)}\right)^2 \varphi
+ \omega \left( \phi^{(0)} \right) \dot\phi^{(0)} \dot\varphi + V' \left( \phi^{(0)} \right) \varphi \right\} \nn
=& 0 \, , \\
\label{GijS}
\delta G^i_{\ j} =& \epsilon^{klm}\left( \delta^i_{\ k} g^{(0)}_{jn} + \delta^i_{\ n} g^{(0)}_{jk} \right)
\left[ - \left\{ \left( 4H \dot{U} +  2 \ddot{U} \right) \partial_l \partial_t {h^{(S)}}^n_{\ m}
+ 2 \dot{U} \partial_l \partial_t^2 {h^{(S)}}^n_{\ m} \right\} \right.  \nn
& \left. + 2 \dot{U} \left( \partial^k \partial_k \partial_l {h^{(S)}}^n_{\ m}
 - \partial_l \partial_k \partial^n {h^{(S)}}^k_{\ m} \right) \right] \nn
& +F' \left\{ H\left[ \frac{3}{2} \partial_t h \delta^i_{\ j}
 - \frac{3}{2} \partial_t {h^{(S)}}^i_{\ j} \right] \right.  \nn
& - \frac{1}{2} \partial^2_t {h^{(S)}}^i_{\ j} + \left( \frac{1}{2} \partial^2_t h
 - \partial^k \partial_k h
+ \frac{2}{3} \left( \partial^k \partial_k \right)^2 B + \frac{1}{3} \partial^k \partial_k h \right) \delta^i_{\ j}
+ \frac{1}{2} \partial^k \partial_k {h^{(S)}}^i_{\ j}
+ \frac{1}{2} \partial^i \partial_j h \nn
& \left. - \frac{1}{2} \left( \partial^i \partial_k {h^{(S)}}^k_{\ j}
+ g^{(0)im} g^{(0)}_{jl} \partial^l \partial_k {h^{(S)}}^k_{\ m} \right) \right\}  \nn
& +F'' \left\{ \ddot{H} \left[  -42 H {h^{(S)}}^i_{\ j} + \left( - 12 H h
+ \partial_{t} h \right) \delta^i_{\ j}
+ 3 \partial_t {h^{(S)}}^i_{\ j} \right] - 24 {\dot{H}}^2 24 {h^{(S)}}^i_{\ j} \right. \nn
& + \dot{H} \left[ - 72 H^2 {h^{(S)}}^i_{\ j} + 12 \partial_{t} {h^{(S)}}^i_{\ j}
+ \left( - 48 H^2 h - 8 H \partial_{t} h \right) \delta^i_{\ j}
+ \left( 7 \partial_t^2 h + 2 \partial^k \partial_k h
 - 2 \left( \partial^k \partial_k \right)^2 B \right) \delta^i_{\ j}
\right] \nn
& - 12 H^3 h \delta^i_{\ j}
+ H^2 \left[ 5 \partial^2_t h + 2 \partial^k \partial_k h
 - 2 \left( \partial^k \partial_k \right)^2 B \right] \delta^i_{\ j} \nn
& +H\left[ 4 \partial^i \partial_j \partial_t h
+ \left( 6 h - 2 \partial^k \partial_k \partial_t h
 - 2 \left( \partial_t + 4 H \right) \left( \frac{2}{3} \left( \partial^k \partial_k \right)^2 B
+ \frac{1}{3} \partial^k \partial_k h \right) \right) \delta^i_{\ j} \right] \nn
& + \partial^i \partial_j \left( \partial_t^2 h - \partial^k \partial_k h \right) \nn
& + \left( \partial_t^4 h - 2 \partial^k \partial_k \partial_t^2 h \right. \nn
& \left. \left. + \left( \partial_t + 4 H \right)^2 \left( \frac{2}{3} \left( \partial^k \partial_k \right)^2 B
+ \frac{1}{3} \partial^k \partial_k h \right)
+ \partial^k \partial_k \partial^l \partial_l h - \partial^l \partial_l
\left( \frac{2}{3} \left( \partial^k \partial_k \right)^2 B
+ \frac{1}{3} \partial^k \partial_k h \right) \right) \delta^i_{\ j} \right\} \nn
& +F''' \left\{ \dddot{H} \left[ 24 H \partial_t h
+6\partial_t^2 h + 4 \left( \left( \partial^k \partial_k \right)^2 B
 - \partial^k \partial_k h \right)  \right] \delta^i_{\ j} - 36 {\ddot{H}}^2 {h^{(S)}}^i_{\ j} \right. \nn
& +\ddot{H}\left[ \dot{H} \left( -288 H {h^{(S)}}^i_{\ j} + 48 \partial_{t} h \right) \delta^i_{\ j}
+ 144 H^2 \partial_{t} h \delta^i_{\ j}
+H \left( 84 \partial_t^2 h
+ 8 \left( \partial^k \partial_k \right)^2 B - 8 \partial^k \partial_k h \right) \delta^i_{\ j} \right. \nn
& \left. + \left( 12\partial_t^3 h
 -12 \left( \partial_t + 2H \right) \left( \partial^k \partial_k h
+  \left( \frac{2}{3} \left( \partial^k \partial_k \right)^2 B
+ \frac{1}{3} \partial^k \partial_k h \right) \right) \right) \delta^i_{\ j} \right] \nn
& +{\dot{H}}^2 \left[ - 576 H^2 {h^{(S)}}^i_{\ j} + \left(
288 H \partial_{t} h +24\partial_t^2 h + 16 \left( \partial^k \partial_k \right)^2 B
 - 16 \partial^k \partial_k h \right) \delta^i_{\ j} \right] \nn
& +\dot{H} \left[ 192 H^3 \partial_{t} h + 240 H^2 \partial_t^2 h
+ H^2 \left( - 32 \left( \partial^k \partial_k \right)^2 B + 32 \partial^k \partial_k h \right) \right. \nn
& \left. \left. +48H\partial_t^3 h
+ 32 H \partial_t \left( - \left( \partial^k \partial_k \right)^2 B + \partial^k \partial_k h \right)
\right] \delta^i_{\ j} \right\}  \nn
& + F'''' \left\{ {\ddot{H}}^2 \left[ 144 H h + 36\partial_t^2 h
+ 24 \left( \partial^k \partial_k \right)^2 B - 24 \partial^k \partial_k h
\right] \delta^i_{\ j} \right. \nn
& +\ddot{H}{\dot{H}}\left[ 1152 H^2 h + 288H\partial_t^2 h
+ 192 H \left( \left( \partial^k \partial_k \right)^2 B - \partial^k \partial_k h \right) \right]
\delta^i_{\ j} \nn
& \left. + {\dot{H}}^2 \left[ 2304 H^3 h + 576 H^2 \partial_t^2 h
+ 384 H^2 \left( \left( \partial^k \partial_k \right)^2 B - \partial^k \partial_k h \right)
\right] \delta^i_{\ j} \right\} \nn
& + \frac{1}{2} \left\{ \frac{\omega \left(\phi^{(0)}\right)}{2} \left( \dot\phi^{(0)} \right)^2
 - V \left( \phi^{(0)} \right) \right\} {h^{(S)}}^i_{\ j} \nn
& + \frac{1}{2} \left\{ \frac{\omega' \left( \phi^{(0)} \right)}{2}
\left(\dot\phi^{(0)}\right)^2 \varphi
+ \omega \left( \phi^{(0)} \right) \dot\phi^{(0)} \dot\varphi
 - V' \left( \phi^{(0)} \right) \varphi \right\} \delta^i_{\ j} \nn
=& 0 \, , \\
\label{GtiS}
\delta G^t_{\ i} =& \frac{2 \dot U}{a^2} \epsilon_{ikl}
\partial^k \partial^m \partial_t {h^{(S)}}^l_{\ m}
+ F' \left\{ \frac{1}{2} \partial_i \partial^t h - \frac{1}{2} \partial^k \partial^t h^{(S)}_{ik} \right\} \nn
& +F'' \left\{ -4 \dot{H} \partial_i \partial_t 4 h
+ 4 H^2 \partial_i \partial_t h
+ H \left[ - 3 \partial_i \partial_t^2 h
+ 2 \left( \left( \partial^k \partial_k \right)^2 B - \partial^k \partial_k h \right) \right] \right. \nn
& \left. - \partial_i \partial_t^3 h
+ \frac{2}{3}\partial_i \left( \partial_t + 2 H \right) \partial^k \partial_k \partial_t
\left( \left( \partial^k \partial_k \right)^2 B - \partial^k \partial_k h \right) \right\} \nn
& + F''' \left\{ \ddot{H} \left[ - 24 H\partial_i \partial_t h^k_{\ k}
 - 6 \partial_i \partial_t^2 h
 - 4 \partial_i \partial^k \left( \left( \partial^k \partial_k \right)^2 B
+ \partial^k \partial_k h \right) \right] \right. \nn
& \left. + \dot{H} \left[ - 96 H^2 h
 - 24 H \partial_i \partial_t^2 h
+ 16 H \left( \left( \partial^k \partial_k \right)^2 B - \partial^k \partial_k h \right) \right] \right\} \nn
& - \frac{\omega \left(\phi^{(0)}  \right)}{2}\dot{\phi^{(0)}} \partial_i \varphi \nn
=& 0\, ,
\end{align}
where $h^{(S)}_{ij}$ is equal to,
\begin{equation}
\label{decomposeS}
h^{(S)}_{ij} = \frac{1}{3} g^{(0)}_{ij} h
+ \partial_i \partial_j B  - \frac{1}{3} g^{(0)}_{ij} \partial^k \partial_k B \, .
\end{equation}
Eq.~(\ref{FCS5SpEx}) takes the following form,
\begin{align}
\label{FCS5SpEx2}
0 =& \frac{1}{2} h \left( \partial_t + 3 H \right) \left( \omega\left(\phi^{(0)} \right) \dot\phi^{(0)} \right)
 - \frac{1}{2} \left( \partial_t + 3 H \right) \left( h \omega\left(\phi^{(0)} \right) \dot\phi^{(0)} \right) \nn
& - \left( \partial_t + 3 H \right) \left( \omega'\left(\phi^{(0)} \right)
\varphi \dot\phi^{(0)} \right)
 - \left( \partial_t + 3 H \right) \left( \omega\left(\phi^{(0)} \right) \dot\varphi \right)
+ a^3 \omega\left(\phi^{(0)} \right) \partial^k \partial_k \varphi
 - V''\left(\phi^{(0)} \right) \varphi \, .
\end{align}
Since the Chern-Simons term does not contribute to the above
equations, as in the $F(R)$ gravity with a scalar field, there
appear two propagating scalar modes. It is notable, and expected
though, that the Chern-Simons term affects solely the tensor
gravitational wave modes, and not the spin-0 mode.

\section{Summary}

In summary, we have investigated the gravitational wave in the
context of Chern-Simons axion $F(R)$ gravity. For the spin-0
scalar mode and the spin-1 vector mode, we demonstrated that for
these modes, the situation is not changed from the standard $F(R)$
gravity coupled with quintessence type scalar field, and in
addition, the Chern-Simons axion term does not affect the
propagation of these modes. This result was also known for the
case of Chern-Simons axion Einstein gravity, as it was shown in
Ref.~\cite{Hwang:2005hb}. Actually, the Chern-Simons term does not
affect the scalar perturbations at all, and it affects solely the
tensor perturbations. As a result, we have two propagating scalar
modes and no propagating vector mode. With regard to the
propagation of the spin-2 graviton mode, the structure of the
dispersion relations become more complicated in comparison with
the Einstein gravity with the Chern-Simons axion, but the
qualitative features of the propagating modes are not changed, and
actually non-equivalent polarization modes occur in both Einstein
Chern-Simons and $F(R)$ gravity Chern-Simons theory. Our study is
focused mainly on primordial gravitational modes, so in a future
work we shall address several related issues, such as the
conservation of the amplitude of gravitational waves at large
scales, and the effect of the function $U(\phi)$ and of the $F(R)$
gravity itself on the polarization asymmetry of the primordial
gravitational waves.

Although the dispersion relation (\ref{GijS6}) tells that the
qualitative structure of the tensor modes is not extensively
changed in comparison to that corresponding to the Chern-Simons
axion Einstein gravity, there appear rather strange behaviors in
the very high energy mode where $k\gg H$ and $k \gg \frac{F'}{\dot
U}$. In the mode the frequency $\omega$ is always complex and
proportional to the square root of the wave number $k$. Therefore
an amplified and a decaying  gravitational wave mode always occur,
and also both the group velocity and the phase velocity are
proportional to $\frac{1}{\sqrt{k}}$. Then both the group velocity
and the propagating velocity become smaller for larger $k$, and
even vanish in the limit of $k\to \infty$.

Finally let us note that in the present work we have found two
propagating scalar modes, with the one being the scalar mode which
appears commonly in the context of higher derivative gravity
\cite{Bogdanos:2009tn}, and with the other being the pseudo-scalar
mode corresponding to the axion scalar. Usually the scalar mode
does not mix with the pseudo-scalar mode, but if the parity
symmetry is broken by the non-trivial value of the Chern-Simons
term, a mixing can occur in general.

\section*{Acknowledgments}
This work is supported by MINECO (Spain), FIS2016-76363-P, and by
project 2017 SGR247 (AGAUR, Catalonia) (S.D.O). This work is also
supported by MEXT KAKENHI Grant-in-Aid for Scientific Research on
Innovative Areas ``Cosmic Acceleration'' No. 15H05890 (S.N.) and
the JSPS Grant-in-Aid for Scientific Research (C) No. 18K03615
(S.N.). The work of A.P. is performed according to the Russian
Government Program of Competitive Growth of Kazan Federal
University. The work of A.P. was also supported by the Russian
Foundation for Basic Research Grant No 19-02-00496.

\section*{Appendix: Detailed Form of Perturbed Einstein Tensor Components}

In this Appendix we present the detailed form of the tensor
expressions needed in the text, but have quite extended form. The
full expression of the perturbed Einstein tensor is,
\begin{align} \label{FCS15} \delta G_{\mu \nu}=
& \frac{1}{2} F\left( R^{(0)} \right) h_{\mu\nu}
 - \frac{1}{2}\Big(\nabla^{(0)}_\mu
\nabla^{(0)\, \rho} h_{\nu\rho} + \nabla^{(0)}_\nu \nabla^{(0)\,
\rho} h_{\mu\rho} - \Box^{(0)} h_{\mu\nu}
 - \nabla^{(0)}_\mu \nabla^{(0)}_\nu \left(g^{(0)\, \rho\lambda} h_{\rho\lambda}\right) \nn
& - 2R^{(0)\, \lambda\ \rho}_{\ \ \ \ \ \nu\ \mu} h_{\lambda\rho}
+ R^{(0)\, \rho}_{\ \ \ \ \ \mu} h_{\rho\nu} + R^{(0)\, \rho}_{\ \
\ \ \ \nu} h_{\rho\mu} \Big) F' \left( R^{(0)} \right) \nn & +
\frac{1}{2}g^{(0)}_{\mu\nu} F' \left( R^{(0)} \right) \left( -
h_{\rho\sigma} R^{(0)\, \rho\sigma} + \nabla^{(0)\, \rho}
\nabla^{(0)\, \sigma} h_{\rho\sigma}
 - \Box^{(0)} \left(g^{(0)\, \rho\sigma} h_{\rho\sigma}\right) \right) \nn
& + \left( - R^{(0)} _{\mu\nu} + \nabla^{(0)}_\mu \nabla^{(0)}_\nu
 - g^{(0)}_{\mu\nu} \Box^{(0)} \right) \left( F'' \left( R^{(0)} \right)
\left( - h_{\rho\sigma} R^{(0)\, \rho\sigma} + \nabla^{(0)\, \rho}
\nabla^{(0)\, \sigma} h_{\rho\sigma}
 - \Box^{(0)} \left(g^{(0)\, \rho\sigma} h_{\rho\sigma}\right) \right) \right) \nn
& + \frac{1}{2}g^{(0) \kappa\lambda}\left( \nabla^{(0)}_\mu
h_{\nu\lambda} + \nabla^{(0)}_\nu h_{\mu\lambda}
 - \nabla^{0}_\lambda h_{\mu\nu} \right) \partial_\kappa F' \left( R^{(0)} \right) \nn
& + g^{(0)}_{\mu\nu} g^{(0)\rho\tau} g^{(0)\sigma\eta}
h_{\rho\sigma} \nabla^{(0)}_\tau \nabla^{(0)}\eta F' \left(
R^{(0)} \right) - \frac{1}{2} g^{(0)}_{\mu\nu} g^{(0)\rho\sigma}
g^{(0) \kappa\lambda}\left( \nabla^{(0)}_\rho h_{\sigma\lambda} +
\nabla^{(0)}_\sigma h_{\rho\lambda}
 - \nabla^{0}_\lambda h_{\rho\sigma} \right) \partial_\kappa F' \left( R^{(0)} \right) \nn
& + \frac{1}{2} \left(
 - \frac{\omega \left(\phi^{(0)} \right)}{2} g^{(0)\rho\sigma}\partial_\rho \phi^{(0)}
\partial_\sigma \phi^{(0)}  - V\left( \phi^{(0)} \right) \right) h_{\mu\nu}
+ \frac{\omega\left( \phi^{(0)} \right)}{4}g^{(0)}_{\mu\nu}
h_{\rho\sigma}
\partial^\rho \phi^{(0)} \partial^\sigma \phi^{(0)} \nn
& + 2 \left( h_{\mu\xi} g^{(0)}_{\nu\sigma} + h_{\mu\sigma}
g^{(0)}_{\nu\xi} + g^{(0)}_{\mu\xi} h_{\nu\sigma} +
g^{(0)}_{\mu\sigma} h_{\nu\xi} \right) \tilde\epsilon^{(0)\,
\zeta\eta\rho\xi} \nabla^{(0)}_\tau \nabla^{(0)}_\rho \left( U
\left(\phi^{(0)} \right) R^{(0)\, \tau\sigma}_{\ \ \ \ \ \
\zeta\eta} \right) \nn
& + 2 \left( g^{(0)}_{\mu\xi} g^{(0)}_{\nu\sigma} +
g^{(0)}_{\mu\sigma} g^{(0)}_{\nu\xi} \right) \left\{
 - \frac{1}{2} g^{(0)\, \alpha\beta} h_{\alpha\beta} \tilde\epsilon^{(0)\, \zeta\eta\rho\xi}
\nabla^{(0)}_\tau \nabla^{(0)}_\rho \left( U
\left(\phi^{(0)}\right) R^{(0)\, \tau\sigma}_{\ \ \ \ \ \
\zeta\eta} \right) \right. \nn & + \tilde\epsilon^{(0)\,
\zeta\eta\rho\xi} h_{\alpha\beta} \nabla^{(0)\, \alpha}
\nabla^{(0)}_\rho \left( U \left(\phi^{(0)} \right) R^{(0)\,
\sigma\beta}_{\ \ \ \ \ \ \zeta\eta} \right) \nn & - \frac{1}{2}
\tilde\epsilon^{(0)\, \zeta\eta\rho\xi} g^{(0)\, \tau\alpha}
\nabla^{(0)}_\tau \nabla^{(0)}_\rho \left( 2 U \left(\phi^{(0)}
\right) g^{(0)\, \sigma\beta} \left( \nabla^{(0)}_\zeta \left(
\nabla^{(0)}_\eta h_{\alpha\beta} + \nabla^{(0)}_\alpha
h_{\eta\beta} - \nabla^{(0)}_\beta h_{\eta\alpha} \right) \right)
\right) \nn & - \frac{1}{2} \tilde\epsilon^{(0)\,
\zeta\eta\rho\xi} g^{(0)\, \tau\alpha} \nabla^{(0)}_\tau \left( U
\left(\phi^{(0)} \right) \left( g^{(0)\, \sigma\beta} \left(
\nabla^{(0)}_\rho h_{\beta\gamma} + \nabla^{(0)}_\gamma
h_{\rho\beta} - \nabla^{(0)}_\beta h_{\rho\gamma} \right) R^{(0)\,
\gamma}_{\ \ \ \ \ \alpha\zeta\eta} \right. \right. \nn & -
g^{(0)\, \gamma\beta} \left( \nabla^{(0)}_\rho h_{\beta\alpha} +
\nabla^{(0)}_\alpha h_{\rho\beta} - \nabla^{(0)}_\beta
h_{\rho\alpha} \right) R^{(0)\, \sigma}_{\ \ \ \ \
\gamma\zeta\eta}
 - g^{(0)\, \gamma\beta} \left( \nabla^{(0)}_\rho h_{\beta\zeta}
+ \nabla^{(0)}_\zeta h_{\rho\beta} - \nabla^{(0)}_\beta
h_{\rho\zeta} \right) R^{(0)\, \sigma}_{\ \ \ \ \
\alpha\gamma\eta} \nn & \left. \left. - g^{(0)\, \gamma\beta}
\left( \nabla^{(0)}_\rho h_{\beta\eta} + \nabla^{(0)}_\eta
h_{\rho\beta} - \nabla^{(0)}_\beta h g_{\rho\eta} \right) R^{(0)\,
\sigma}_{\ \ \ \ \ \alpha\zeta\gamma} \right) \right) \nn & -
\frac{1}{2} \tilde\epsilon^{(0)\, \zeta\eta\rho\xi} g^{(0)
\tau\alpha} \left( - g^{(0)\ \beta\gamma} \left( \nabla^{(0)}_\tau
h_{\gamma\rho} + \nabla^{(0)}_\rho h_{\tau\gamma} -
\nabla^{(0)}_\gamma h_{\tau\rho} \right) \nabla^{(0)}_\beta \left(
U \left(\phi^{(0)} \right) R^{(0)\, \sigma}_{\ \ \ \ \
\alpha\zeta\eta} \right) \right. \nn & + g^{(0)\, \sigma\gamma}
\left( \nabla^{(0)}_\tau h_{\gamma\beta} + \nabla^{(0)}_\beta
h_{\tau\gamma} - \nabla^{(0)}_\gamma h_{\tau\beta} \right)
\nabla^{(0)}_\rho \left( U \left(\phi^{(0)} \right) R^{(0)\,
\beta}_{\ \ \ \ \ \alpha\zeta\eta} \right) \nn & - g^{(0)\,
\beta\gamma} \left( \nabla^{(0)}_\tau h_{\gamma\alpha} +
\nabla^{(0)}_\alpha h_{\tau\gamma} - \nabla^{(0)}_\gamma
h_{\tau\alpha} \right) \nabla^{(0)}_\rho \left( U \left(
\phi^{(0)} \right) R^{(0)\, \sigma}_{\ \ \ \ \ \beta\zeta\eta}
\right) \nn & - g^{(0)\, \beta\gamma} \left( \nabla^{(0)}_\tau
h_{\gamma\zeta} + \nabla^{(0)}_\zeta h_{\tau\gamma} -
\nabla^{(0)}_\gamma h_{\tau\zeta} \right) \nabla^{(0)}_\rho \left(
U \left( \phi^{(0)} \right) R^{(0)\, \sigma}_{\ \ \ \ \
\alpha\beta\eta} \right) \nn & \left. \left. - g^{(0)\,
\beta\gamma} \left( \nabla^{(0)}_\tau h_{\gamma\eta} +
\nabla^{(0)}_\eta h_{\tau\gamma} - \nabla^{(0)}_\gamma
h_{\tau\eta} \right) \nabla^{(0)}_\rho \left( U \left( \phi^{(0)}
\right) R^{(0)\, \sigma}_{\ \ \ \ \ \alpha\zeta\beta} \right)
\right) \right\} \nn
& + \frac{\omega \left(\phi^{(0)} \right)}{4}\partial_\rho
\phi^{(0)} \partial_\sigma \phi^{(0)} h^{\rho\sigma}
g^{(0)}_{\mu\nu} + \frac{1}{2} \left\{ - \frac{\omega
\left(\phi^{(0)}\right)}{2}
\partial_\mu \phi^{(0)} \partial^\mu \phi^{(0)} - V \left( \phi^{(0)} \right) \right\} h_{\mu\nu} \nn
& + \frac{1}{2} \left\{ - \frac{\omega' \left( \phi^{(0)}
\right)}{2}
\partial_\mu \phi^{(0)} \partial^\mu \phi^{(0)} \varphi
 - \omega \left( \phi^{(0)} \right) \partial_\mu \phi^{(0)} \partial^\mu \varphi
 - V' \left( \phi^{(0)} \right) \varphi \right\} g^{(0)}_{\mu\nu}
+ \frac{\omega' \left(\phi^{(0)} \right)}{2}\partial_\mu
\phi^{(0)} \partial_\nu \phi^{(0)} \varphi \nn & + \frac{\omega
\left(\phi^{(0)}  \right)}{2} \left( \partial_\mu \varphi
\partial_\nu \phi^{(0)} + \partial_\mu \phi^{(0)} \partial_\nu
\varphi \right) + 2 \left( g^{(0)}_{\mu\xi} g^{(0)}_{\nu\sigma} +
g^{(0)}_{\mu\sigma} g^{(0)}_{\nu\xi} \right) \nabla^{(0)}_\tau
\nabla^{(0)}_\rho \left( U' \left( \phi^{(0)} \right) \varphi
\tilde\epsilon^{(0) \zeta\eta\rho\xi} R^{(0) \tau\sigma}_{\ \ \ \
\ \zeta\eta} \right) \nn = & 0 \, .
\end{align}
Moreover, the non-zero components of the perturbed Einstein tensor
are $\delta G^i_{\ j}$,
\begin{align}
\label{Gij} \delta G^i_{\ j} \equiv & g^{(0)i\mu} \delta G_{\mu j}
\nn =& \epsilon^{klm}\left( \delta^i_{\ k} g^{(0)}_{jn} +
\delta^i_{\ n} g^{(0)}_{jk} \right) \left[ - \left\{ \left( 4H
\dot{U} +  2 \ddot{U} \right) \partial_l \partial_t h^n_{\ m} + 2
\dot{U} \partial_l \partial_t^2 h^n_{\ m} \right\} + 2 \ddot{U}
\partial_l \partial^n h^t_{\ m} \right.  \nn & \left. + 2 \dot{U}
\left( - \partial^n \partial_l \partial_t h^t_{\ m} + \partial^k
\partial_k \partial_l h^n_{\ m} - \partial_l \partial_k \partial^n
h^k_{\ m} \right) \right] \nn & +F' \left\{ -\left( \dot{H} +3H^2
\right) h^t_{\ t} \delta^i_{j\ } +H\left[ \left( \partial_t
\left(\frac{3}{2}h^k_{\ k} - h^t_{\ t} \right) + \partial^k h^t_{\
k} \right) \delta^i_{\ j}
 - \frac{3}{2} \partial_t h^i_{\ j} - \frac{1}{2} \partial^i h^t_{\ j}
 - \frac{1}{2} \partial_j h^{ti} \right] \right.  \nn
& - \frac{1}{2} \partial^2_t h^i_{\ j} + \left( \frac{1}{2}
\partial^2_t h^k_{\ k}
 - \partial^k \partial_k \left( h^t_{\ t} + h^l_{\ l} \right)
+ \partial^k \partial^l h_{kl} + 2 \partial^t \partial^k h_{tk}
\right) \delta^i_{\ j} + \frac{1}{2} \partial^k \partial_k h^i_{\
j} + \frac{1}{2} \partial^i \partial_j \left( h^t_{\ t} + h^k_{\
k} \right) \nn & \left. - \frac{1}{2} \left( \partial^i \partial_k
h^k_{\ j} + \partial^t \partial^i h_{tj} + g^{(0)im} g^{(0)}_{jl}
\left( \partial^l \partial_k h^k_{\ m} + \partial^t \partial^l
h_{tm} \right) \right) \right\}  \nn & +F'' \left\{ -12 \,
\dddot{H} \, h^t_{\ t} \delta^i_{\ j} + \ddot{H} \left[  -42 H
h^i_{\ j} + \left( H \left(-33 h^t_{\ t} -12 h^k_{\ k} \right) -12
\partial_{t} h^t_{\ t}  + \partial_{t} h^k_{\ k} - 2 \partial^k
h^t_{\ k} \right) \delta^i_{\ j} + 3 \partial_{t} h^i_{\ j}
\right. \right. \nn & \left. + 3\partial^i h^t_{\ j} + 3\partial_j
h^{ti}\right] +{\dot{H}}^2 \left( -30 h^t_{\ t} \delta^i_{\ j} -24
h^i_{\ j} \right) \nn & + \dot{H} \left[ - 72 H^2 h^i_{\ j} + 12
\partial_{t} h^i_{\ j} + 12 \partial^i h^t_{\ j} + 12 \partial_j
h^{ti} + \left( H^2\left( 42 h^t_{\ t} -48 h^k_{\ k} \right) +
H\left( -51\partial_{t} h^t_{\ t} - 8\partial_{t} h^k_{\ k}
-44\partial^k h^t_{\ k} \right) \right) \delta^i_{\ j} \right. \nn
& \left. - 6 \partial^i \partial_j h^t_{t} + \left( \partial^2_t
\left(-12h^t_{\ t} +7h^k_{\ k} \right)
 + 9 \partial^k \partial_k h^t_{\ t} +2\partial_t \partial^k h^t_{\ k}
+ 3 \partial^k \partial_k h^l_{\ l} - 3 \partial^k \partial^l
h_{kl} \right) \delta^i_{\ j} \right] +36H^4 h^t_{\ t} \delta^i_{\
j} \nn & +H^3\left[ \partial_{t}\left(-15 h^t_{\ t} -12 h^k_{\ k}
\right)
 -12 \partial^k h^t_{\ k} \right] \delta^i_{\ j} \nn
& + H^2 \left[ 12 \partial^i \partial_j h^t_{\ t} + \left(
\partial^2_t \left( -18 h^t_{\ t} +5 h^k_{\ k} \right) +
\partial^k \partial_k \left( 15h^t_{\ t} + 3h^k_{\ k} \right)
 - 3 \partial^k \partial^l h_{kl}
-14 \partial^t \partial^k h^t_{\ k} \right) \delta^i_{\ j} \right]
\nn & +H\left[ \partial^i \partial_j \left(  - 3 \partial_t h^t_{\
t} + 4 \partial_t h^k_{\ k} + 4 h^t_{\ l} \right) \right. \nn &
\left. + \left( \partial_t^3 \left(-3 h^t_{\ t} +6 h^k_{\ k}
\right) + 5 \partial^k \partial_k \partial_t h^t_{\ t}  -2
\partial^k \partial_k \partial_t h^l_{\ l}
 - 2 \partial^k \partial^l \partial_t h_{kl}
 - 4 \partial^k \partial_k \partial^l h^t_{\ l} \right) \delta^i_{\ j} \right] \nn
& + \partial^i \partial_j \left( \partial_t^2 h^k_{\ k} -
\partial^k \partial_k h^t_{\ t}
 - \partial^k \partial_k h^m_{\ m} + 2 \partial^l \partial_t h^t_{\ l} \right) \nn
& + \left( \partial_t^4 h^k_{\ k} +  2\partial^k \partial_t^3
h^t_{\ k}
 - \partial^k \partial_k \partial_t^2 h^t_{\ t}
 - 2 \partial^k \partial_k \partial_t^2 h^l_{\ l} \right. \nn
& \left. \left. + \partial^k \partial^l \partial_t^2 h_{kl} +
\partial^k \partial_k \partial^l \partial_l h^t_{\ t} + \partial^k
\partial_k \partial^l \partial_l h^m_{\ m} - \partial^k \partial_k
\partial^l \partial^m h_{lm}
 -2 \partial^k \partial_k \partial^l \partial_t h^t_{\ l} \right) \delta^i_{\ j} \right\} \nn
& +F''' \left\{ \dddot{H} \left[ - 36 \dot{H} h^t_{\ t} - 72 H^2
h^t_{\ t} + H \left( \partial_t \left( - 18 h^t_{\ t} + 24 h^k_{\
k} \right) + 24 \partial^k h^t_{\ k} \right) \right. \right. \nn &
\left. +6\partial_t^2 h^k_{\ k}
 - 6 \partial^k \partial_k \left(h^t_{\ t} + h^l_{\ l} \right)
+ 6 \partial^k \partial^l h_{kl} + 12 \partial^k \partial_t h^t_{\
k} \right] \delta^i_{\ j} \nn & + {\ddot{H}}^2 \left[ -90h^t_{\ t}
\delta^i_{\ j} -36h^i_{\ j} \right] \nn & +\ddot{H}\left[ \dot{H}
\left( -288 H h^i_{\ j} + \left( -936 H h^t_{\ t} + \partial_{t}
\left(-108h^t_{\ t} +48h^k_{\ k} \right) + 48 \partial^k h^t_{\ k}
\right) \delta^i_{\ j} \right)
 -432H^3 h^t_{\ t}  \delta^i_{\ j} \right. \nn
&  +H^2 \left( \partial_{t} \left(-252h^t_{\ t} +144h^k_{\ k}
\right) + 48 \partial^k h^t_{\ k} \right) \delta^i_{\ j} \nn & +H
\left( \partial_t^2 \left(-36h^t_{\ t} +84 h^k_{\ k} \right)
 - 12\partial^k \partial_k \partial_t \left( h^t_{\ t} + h^k_{\ k} \right) + 12 \partial^k \partial^l h_{kl}
+ 72 \partial^k \partial_t h^t_{\ k} \right) \delta^i_{\ j} \nn &
\left. + \left( 12\partial_t^3 h^k_{\ k} + 24 \partial^k
\partial_t^2 h^t_{\ k}
 -12 \partial^k \partial_k \left( h^t_{\ t} +h^k_{\ k} \right)
+ 12 \partial_k \partial^l \partial_t h^k_{\ l} \right)
\delta^i_{\ j} \right] \nn & +{\dot{H}}^2 \left[ - 576 H^2 h^i_{\
j} + \left\{
 -2016 H^2 h^t_{\ t} +H \left( \partial_{t}
\left( - 504 h^t_{\ t} + 288 h^k_{\ k} \right) \right. \right.
\right. \nn & \left. \left. \left. + 288 \partial^k h^t_{\ k}
\right) +24\partial_t^2 h^k_{\ k}
 - 24 \partial^k \partial_k \left( h^t_{\ t} + h^k_{\ k} \right)
+ 24 \partial^k \partial^l h_{kl} + 48 \partial^k \partial_t
h^t_{\ k} \right) \delta^i_{\ j} \right] \nn & +\dot{H} \left[
-576 H^4 h^t_{\ t} +H^3 \left( \partial_{t} \left( -720 h^t_{\ t}
+192 h^k_{\ k} \right)
 - 192 \partial^k h^t_{\ k} \right) \right. \nn
& +H^2 \partial_t^2 \left(-144h^t_{\ t} +240h^k_{\ k} \right) +
H^2 \left( 48 \partial^k \partial_k \left( h^t_{\ t} + h^k_{\ k}
\right)
 - 48 \partial^k \partial^l h_{kl} + 96 \partial^k \partial_t h^t_{\ k} \right) \nn
& \left. \left. +48H\partial_t^3 h^k_{\ k} + H \left( 96
\partial^k \partial_t^2 h^t_{\ k}
 -48 \partial^k \partial_k \partial_t \left( h^t_{\ t} + h^k_{\ k} \right)
+ 48 \partial^k \partial^l \partial_t h_{kl} \right) \right]
\delta^i_{\ j} \right\}  \nn & + F'''' \left\{ {\ddot{H}}^2 \left[
-216 \dot{H} h^t_{\ t} -432 H^2 h^t_{\ t} +H \left( \partial_{t}
(-108h^t_{\ t} +144 h^k_{\ k} \right) \right. \right. \nn & \left.
\left. + 144 \partial^k h^t_{\ k} \right) + 36\partial_t^2 h^k_{\
k}
 -36 \partial^k \partial_k \left( h^t_{\ t} +h^k_{\ k} \right)
+ 36 \partial^k \partial^l h_{kl} + 72 \partial^k \partial_t
h^t_{\ k} \right] \delta^i_{\ j} \nn & -1728\ddot{H}{\dot{H}}^2
h^t_{\ t} \delta^i_{\ j} +\ddot{H}{\dot{H}}\left[ -3456 H^3 h^t_{\
t} + H^2\left( \partial_{t} \left( -864 h^t_{\ t} +1152 h^k_{\ k}
\right) + 1152 \partial^k h^t_{\ k} \right) \right. \nn & \left.
+288H\partial_t^2 h^k_{\ k} + 288 H \left( - \partial^k \partial_k
\left( h^t_{\ t} + h^l_{\ l} \right) + \partial^k \partial^l
h_{kl} + 2 \partial^k \partial_t h^t_{\ k} \right) \right]
\delta^i_{\ j}  - 3456{\dot{H}}^3 H^2 h^t_{\ t} \delta^i_{\ j} \nn
& + {\dot{H}}^2 \left[-6912  H^4 h^t_{\ t} +H^3 \left(
\partial_{t} \left( -1728 h^t_{\ t} +2304 h^k_{\ k} \right) + 2304
\partial^k h^t_{\ k} \right) \right. \nn & \left. \left. +576 H^2
\partial_t^2 h^k_{\ k} + 576 H^2 \left( - \partial^k \partial_k
\left(h^t_{\ t} + h^l_{\ l} \right) + \partial^k \partial^l h_{kl}
+ 2 \partial^k \partial_t h^t_{\ k} \right) \right] \delta^i_{\ j}
\right\} \nn & + \frac{\omega \left(\phi^{(0)} \right)}{4} \left(
\dot\phi^{(0)} \right)^2 h^{tt} \delta^i_{\ j} + \frac{1}{2}
\left\{ \frac{\omega \left(\phi^{(0)}\right)}{2} \left(
\dot\phi^{(0)} \right)^2
 - V \left( \phi^{(0)} \right) \right\} h^i_{\ j} \nn
& + \frac{1}{2} \left\{ \frac{\omega' \left( \phi^{(0)}
\right)}{2} \left(\dot\phi^{(0)}\right)^2 \varphi + \omega \left(
\phi^{(0)} \right) \dot\phi^{(0)} \dot\varphi
 - V' \left( \phi^{(0)} \right) \varphi \right\} \delta^i_{\ j} \nn
=& 0 \, .
\end{align}

\end{document}